\documentclass[copyright]{eptcs}

\providecommand{\event}{QAPL 2015}

\providecommand{\firstpage}{1}

\usepackage{amsmath,
	    amsfonts,
	    amstext,
	    amssymb,
	    stmaryrd,
	    latexsym,
	    comment,
	    graphicx}

\def\ms#1{\null\ifmmode\mathord{\mathcode`-="702D\it #1\mathcode`\-="2200}%
	\else$\mathord{\mathcode`-="702D\it #1\mathcode`\-="2200}$\fi}

\newcommand{\cws}[2]
	{\\ \centerline{$#2$} \\[-#1pt]}

\newlength{\spacelen}

\newcommand{\bibtrick}[1]
	{}

\newcommand{\lsp}
	{[ \! [}
\newcommand{\rsp}
	{] \! ]}

\newcommand{\calb}
        {\mathcal{B}}

\newcommand{\calm}
        {\mathcal{M}}

\newcommand{\calt}
        {\mathcal{T}}

\newcommand{\natns}
	{\mathbb{N}}

\newcommand{\realns}
	{\mathbb{R}}

\newcommand{\arrow}[2]
        {\, {\auxarrow\limits^{#1}}_{#2} \,}
\newcommand{\auxarrow}
	{\mathop{- \!\!\!\!\! \longrightarrow}}

\newcommand{\warrow}[2]
        {\, {\wauxarrow\limits^{#1}}_{#2} \,}
\newcommand{\wauxarrow}
	{\mathop{= \!\!\!\! \Longrightarrow}}

\newcommand{\tarrow}[2]
        {\, {\tauxarrow\limits^{#1}}_{#2} \,}
\newcommand{\tauxarrow}
	{\mathop{- \!\!\!\! - \!\!\! \leadsto}}

\newcommand{\subtarrow}[2]
        {\, {\subtauxarrow\limits^{#1}}_{#2} \,}
\newcommand{\subtauxarrow}
	{\mathop{- \! \leadsto}}

\newcommand{\sbis}[1]
	{\sim_{#1}}

\newcommand{\wbis}[1]
        {\approx_{#1}}

\newcommand{\pco}[1]
	{\mathop{\Vert_{#1}}}

\newcommand{\fullbox}
	{{\mbox{}\nolinebreak\hfill{$\rule{2mm}{2mm}$}}}

\newtheorem{new_theorem}
	{Theorem}[section]

\newtheorem{new_definition}
	[new_theorem]{Definition}

\newtheorem{new_remark}
	[new_theorem]{Remark}

\newtheorem{new_example}
	[new_theorem]{Example}

\newtheorem{new_lemma}
	[new_theorem]{Lemma}

\newtheorem{new_proposition}
	[new_theorem]{Proposition}

\newtheorem{new_corollary}
	[new_theorem]{Corollary}

\newenvironment{definition}
	{\begin{new_definition}\rm}
	{\end{new_definition}}

\newenvironment{example}
	{\begin{new_example}\rm}
	{\end{new_example}}

\newenvironment{theorem}
	{\begin{new_theorem}\rm}
	{\end{new_theorem}}

\begin{document}

\title{Expected-Delay-Summing Weak Bisimilarity \\
       for Markov Automata}

\author{Alessandro Aldini \qquad Marco Bernardo
\institute{Dipartimento di Scienze di Base e Fondamenti, Universit\`a di Urbino, Italy}}

\def\titlerunning{Expected-Delay-Summing Weak Bisimilarity for Markov Automata}
\def\authorrunning{A.~Aldini, M.~Bernardo}

\maketitle


\begin{abstract}

A new weak bisimulation semantics is defined for Markov automata that, in addition to abstracting from
internal actions, sums up the expected values of consecutive exponentially distributed delays possibly
intertwined with internal actions. The resulting equivalence is shown to be a congruence with respect to
parallel composition for Markov automata. Moreover, it turns out to be comparable with weak bisimilarity for
timed labeled transition systems, thus constituting a step towards reconciling the semantics for stochastic
time and deterministic time.

\end{abstract}

%
%
\section{Introduction}
\label{sec:intro}
%
%

Markov automata~\cite{EHZ10} integrate Segala's simple probabilistic automata~\cite{Seg95a} and Hermanns'
interactive Markov chains~\cite{Her02}, thus resulting in very expressive models. Markov automata feature
two types of transitions, one for \emph{action execution} and one for \emph{time passing}. The choice among
the actions enabled in a state is nondeterministic, the execution of the selected action is instantaneous,
and the reached state is established according to a probability distribution. Time passing is described by
means of exponentially distributed delays governed by the race policy, with the execution of internal
actions taking precedence over such delays so to enforce maximal progress.

Markov automata come equipped with two compositional semantics, respectively based on \emph{strong} and
\emph{weak} bisimilarity. While the former is the obvious combination of strong bisimilarity for
probabilistic automata and strong bisimilarity for interactive Markov chains, the latter is more complicated
due to certain desirable identifications that should be achieved when abstracting from internal actions.
This has been accomplished by suitably defining weak bisimilarity over state probability subdistributions
rather than over individual states~\cite{EHZ10}. The resulting equivalence has been shown to provide a sound
and complete proof methodology for a touchstone equivalence called reduction barbed congruence~\cite{DH13}.

Weak bisimilarity for Markov automata abstracts from \emph{internal instantaneous actions}. However, in the
setting of labeled transition systems enriched with deterministic delays, which are at the basis of models
such as timed automata~\cite{AD94}, the weak bisimulation semantics appeared in the literature (e.g.,
\cite{Yi91,MT92,AM96}) are also capable of abstracting from \emph{sequences of delays} possibly intertwined
with internal actions, in the sense that those delays can be reduced to a single one equal to the \emph{sum}
of the original delays. Some work in this direction has recently been done for Markovian process calculi
with durational actions that, unlike Markov automata, feature neither nondeterminism nor probabilistic
branching. \linebreak More precisely, in~\cite{Ber15} a weak bisimilarity has been proposed, which is
capable of abstracting from internal actions that are exponentially timed by summing up their \emph{expected
delays}.

	\begin{figure}[t]

\centerline{\includegraphics[width=4.7in]{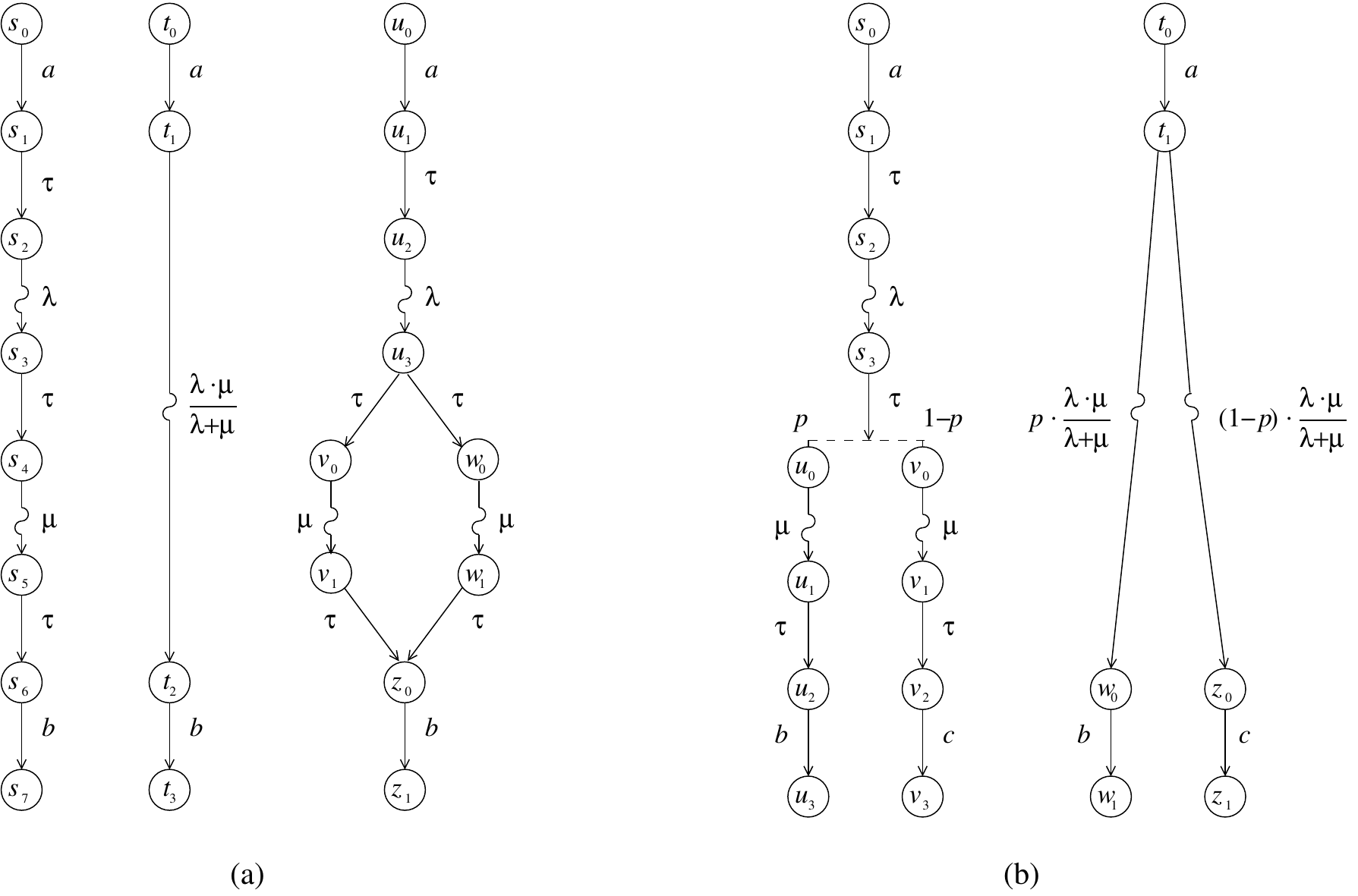}}
\caption{Merging $\tau$-transitions, timed transitions, and nondeterministic choices}
\label{fig:tau_exp_nondet_choice}

	\end{figure}

The purpose of this paper is to study an \emph{expected-delay-summing} weak bisimulation semantics in the
more expressive setting of Markov automata. As we will see, defining such a semantics is a challenging task
due to the need of balancing disparate demands related to nondeterministic, probabilistic, and timing
behaviors. Additionally, in~\cite{Ber15} a tradeoff has emerged between compositionality, i.e., being a
congruence with respect to parallel composition, and pseudo-aggregation exactness, i.e., preserving
stationary-state reward-based performance measures, in the sense that in the Markovian setting it is not
possible to define an expected-delay-summing weak equivalence that enjoys both properties. These facts make
it far from trivial to embody the expected-delay-summing capability of the weak semantics of~\cite{Ber15}
into the weak semantics originally developed for Markov automata in~\cite{EHZ10}.

To clarify the additional identifications that we would like to obtain with respect to~\cite{EHZ10}, let us
consider a few illustrative examples. In these examples, we will depict states as circles, action
transitions as arrows labeled with $a, b, c$ for visible actions and $\tau$ for internal actions, timed
transitions as arrows labeled with positive real rates $\lambda, \mu$ representing the inverses of expected
delays, and probability distributions as dashed lines connecting states.

The first Markov automaton in Fig.~\ref{fig:tau_exp_nondet_choice}(a) has an initial $a$-transition and a
final $b$-transition, with a sequence of timed and $\tau$-transitions in between. All the transitions in the
sequence are reduced to a single timed transition in the second Markov automaton, whose rate has been
computed as the inverse of the expected duration of the entire sequence: $(\frac{1}{\lambda} +
\frac{1}{\mu})^{-1} = \frac{\lambda \cdot \mu}{\lambda + \mu}$. Also the third Markov automaton can be
reduced to the second one, as both branches of the internal nondeterministic choice at state $u_{3}$ have
the same expected duration $\frac{1}{\mu}$.

The two Markov automata in Fig.~\ref{fig:tau_exp_nondet_choice}(b) can be identified as well. The timed and
$\tau$-transitions preceding and following the internal probabilistic choice after state $s_{3}$ can be
reduced to two alternative timed transitions, whose basic rate $\frac{\lambda \cdot \mu}{\lambda + \mu}$ is
respectively multiplied by $p$ and $1 - p$.

	\begin{figure}[t]

\centerline{\includegraphics[width=4.7in]{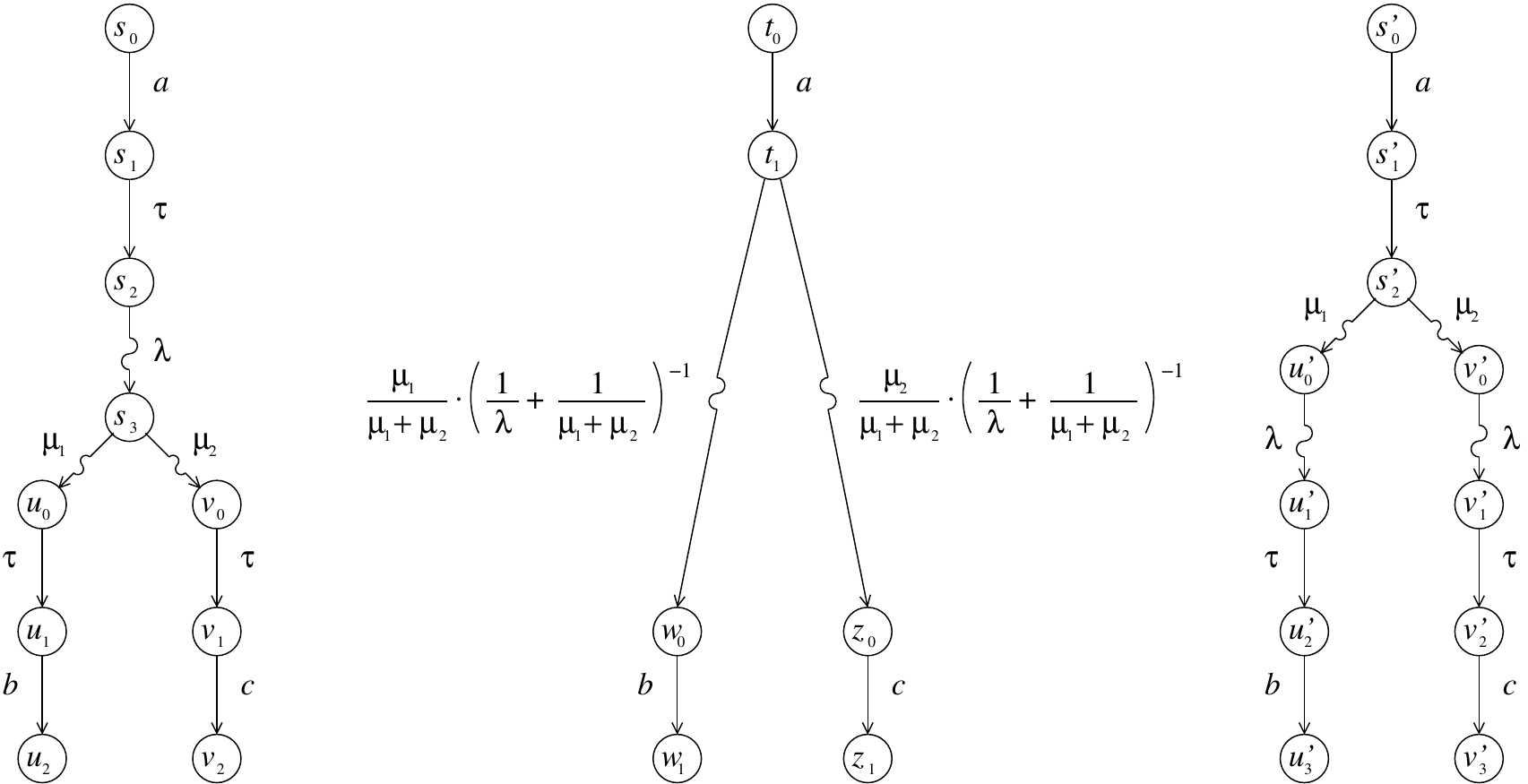}}
\caption{Merging $\tau$-transitions, timed transitions, and stochastic choices}
\label{fig:tau_exp_stoc_choice}

	\end{figure}

In Fig.~\ref{fig:tau_exp_stoc_choice}, the focus is on stochastic choices governed by the race policy,
according to which the execution probability of a timed transition is proportional to its rate. The first
Markov automaton has a stochastic choice at state $s_{3}$ between a $\mu_{1}$-transition and a
$\mu_{2}$-transition, whilst the third Markov automaton has the same stochastic choice at state $s'_{2}$.
Both automata can be seen as equivalent to the second one, in which the stochastic choice and the timed and
$\tau$-transitions preceding and following it are reduced to two alternative timed transitions, whose basic
rate $(\frac{1}{\lambda} + \frac{1}{\mu_{1} + \mu_{2}})^{-1}$ is respectively multiplied by
$\frac{\mu_{1}}{\mu_{1} + \mu_{2}}$ -- probability of taking the left branch -- and $\frac{\mu_{2}}{\mu_{1}
+ \mu_{2}}$ -- probability of taking the right branch.

The paper is organized as follows. In Sect.~\ref{sec:background}, we present some background material and
recall the definition of weak bisimilarity for Markov automata provided in~\cite{EHZ10}. In
Sect.~\ref{sec:eds_weak_bisim}, we introduce our expected-delay-summing weak bisimilarity for Markov
automata by extending the existing one with the construction of~\cite{Ber15}, we prove that it is a
congruence with respect to parallel composition, and we compare its discriminating power with that of the
weak bisimilarity of~\cite{EHZ10}. In Sect.~\ref{sec:det_time}, we propose a comparison between our relation
and the weak bisimilarity for timed labeled transition systems defined in~\cite{Yi91,MT92}, as both
equivalences are capable of summing up expected delays while abstracting from internal actions. Finally, in
Sect.~\ref{sec:concl} we provide some concluding remarks.

%
%
\section{Background}
\label{sec:background}
%
%

%
\subsection{Discrete Probability Subdistributions}
\label{sec:disc_prob_subdistr}
%

Let $\Delta$ be a function from a nonempty, at most countable set $S$ to $\realns_{[0, 1]}$. The support of
$\Delta$ is defined as $\ms{supp}(\Delta) = \{ s \in S \mid \Delta(s) > 0 \}$, while the size of $\Delta$ is
defined as $\ms{size}(\Delta) = \Delta(S)$ where, in general, $\Delta(S') = \sum_{s' \in S'} \Delta(s')$ for
all $S' \subseteq S$.

Function $\Delta$ is a \emph{discrete probability (sub)distribution} over~$S$ iff $\ms{size}(\Delta) = 1$
(resp.\ $\ms{size}(\Delta) \le 1$). \linebreak We denote by $\ms{Subdistr}(S)$ and $\ms{Distr}(S)$ the sets
of subdistributions and distributions over $S$. \linebreak Furthermore, we indicate with $\delta_{s}$ the
Dirac distribution for $s \in S$, i.e., $\delta_{s}(s) = 1$ and $\delta_{s}(s') = 0$ for all $s' \in S
\setminus \{ s \}$.

Given $x \in \realns_{\ge 0}$ and $\Delta \in \ms{Subdistr}(S)$, we denote by $x \odot \Delta$ the function
defined by $(x \odot \Delta)(s) = x \cdot \Delta(s)$ for all $s \in S$. Given $\Delta_{1}, \Delta_{2} \in
\ms{Subdistr}(S)$, we denote by $\Delta_{1} \oplus \Delta_{2}$ the function defined by $(\Delta_{1} \oplus
\Delta_{2})(s) = \Delta_{1}(s) + \Delta_{2}(s)$. These functions are subdistributions when their size does
not exceed $1$. Moreover, given $\Delta_{1} \in \ms{Subdistr}(S_{1})$ and $\Delta_{2} \in
\ms{Subdistr}(S_{2})$, we denote by $\Delta_{1} \otimes \Delta_{2}$ the subdistribution over $S_{1} \times
S_{2}$ defined by $(\Delta_{1} \otimes \Delta_{2})(s_{1}, s_{2}) = \Delta_{1}(s_{1}) \cdot
\Delta_{2}(s_{2})$.

A subdistribution $\Delta$ over $S$ can be viewed as a subset of $S \times \realns_{]0, 1]}$, in which only
elements of $\ms{supp}(\Delta)$ occur, each once with the corresponding probability. In other words,
subdistribution $\Delta$ may be written as $\lsp (s, p) \mid s \in \ms{supp}(\Delta) \land p = \Delta(s)
\rsp$. We also denote by $\Delta \ominus s$ the subdistribution that is obtained from $\Delta$ by removing
the pair $(s, \Delta(s))$ when $s \in \ms{supp}(\Delta)$.

%
\subsection{Markov Automata}
\label{sec:ma}
%

Markov automata~\cite{EHZ10} have two distinct types of transitions: \emph{action transitions} and
\emph{Markov timed \linebreak transitions}. The choice among the action transitions departing from a given
state is nondeterministic. This choice can be influenced by the external environment, except for transitions
labeled with the internal action $\tau$. Once an action transition is chosen, the next state is internally
selected according to some probability distribution, as in probabilistic automata~\cite{Seg95a}.

Each Markov timed transition is labeled with a real number called \emph{rate}, which uniquely identifies an
exponentially distributed delay. As with interactive Markov chains~\cite{Her02}, the choice among the Markov
timed transitions departing from a given state is governed by the \emph{race policy}, which means that the
Markov timed transition that is executed is the one sampling the least duration. Therefore, the execution
probability of a Markov timed transition is proportional to its rate, and the sojourn time associated with a
state having outgoing Markov timed transitions is exponentially distributed with rate given by the sum of
the rates of those transitions.

Different from~\cite{EHZ10}, where Markov automata were introduced for the first time, in the definition
below we explicitly build some assumptions into the model.

	\begin{definition}\label{def:ma}

A \emph{Markov automaton (MA)} is a tuple $(S, A, \! \arrow{}{} \!, \! \tarrow{}{} \!)$ where:

		\begin{itemize}

\item $S$ is a nonempty, at most countable set of states.

\item $A$ is a set of actions containing at least the internal action $\tau$.

\item $\! \arrow{}{} \! \subseteq S \times A \times \ms{Distr}(S)$ is an action-transition relation.

\item $\! \tarrow{}{} \! \subseteq S \times \realns_{\ge 0} \times S$ is a time-transition relation such
that for all $s \in S$:

			\begin{itemize}

\item If $s \tarrow{0}{} s'$ for some $s' \in S$, then $s' = s$ \emph{(zero speed)}.

\item $\sum_{(s, \lambda, s') \in \! \subtarrow{}{}} \lambda < \infty$ \emph{(speed boundedness)}.

\item If $s \arrow{\tau}{} \Delta$ with $\Delta \in \ms{Distr}(S)$, then $s \hspace{0.3cm} / \hspace{-0.5cm}
\tarrow{\lambda}{} s'$ for all $\lambda \in \realns_{\ge 0}$ and $s' \in S$ \emph{(maximal progress)}.
\fullbox

			\end{itemize}

		\end{itemize}

	\end{definition}

Also the notion of parallel composition, although equivalent to the one in~\cite{EHZ10}, is formulated in a
slightly different way. We recall that the first of the three conditions below about the time-transition
relation ensures that the parallel composition of two Markov timed selfloops, each having the same
rate~$\lambda$, results in a Markov timed selfloop with rate $\lambda + \lambda$, as established by the race
policy, instead of~$\lambda$. In the following, $s \hspace{0.3cm} / \hspace{-0.5cm} \arrow{}{}$ and $s
\hspace{0.3cm} / \hspace{-0.5cm} \tarrow{}{}$ stand for the absence of action transitions and Markov timed
transitions, respectively, out of state $s$; an action or rate decoration of the negative arrow means the
absence of transitions labeled with that action or rate.

	\begin{definition}\label{def:par_comp}

Let $\calm_{k} = (S_{k}, A_{k}, \! \arrow{}{k} \!, \! \tarrow{}{k} \!)$ be an MA with initial state $s_{0,
k}$ for $k = 1, 2$. The \emph{parallel composition} of $\calm_{1}$ and $\calm_{2}$ with respect to an action
set $A \subseteq (A_{1} \cup A_{2}) \setminus \{ \tau \}$ is the MA $\calm_{1} \pco{A} \calm_{2} = (S_{1}
\times S_{2}, A_{1} \cup A_{2}, \! \arrow{}{} \!, \! \tarrow{}{} \!)$ with initial state $(s_{0, 1}, s_{0,
2})$ such that:

		\begin{itemize}

\item $(s_{1}, s_{2}) \arrow{a}{} \Delta$ iff one of the following conditions is fulfilled:

			\begin{itemize}

\item $a \in A$, $s_{1} \arrow{a}{1} \Delta_{1}$, $s_{2} \arrow{a}{2} \Delta_{2}$, and $\Delta = \Delta_{1}
\otimes \Delta_{2}$.

\item $a \notin A$, $s_{1} \arrow{a}{1} \Delta_{1}$, and $\Delta = \Delta_{1} \otimes \delta_{s_{2}}$.

\item $a \notin A$, $s_{2} \arrow{a}{2} \Delta_{2}$, and $\Delta = \delta_{s_{1}} \otimes \Delta_{2}$.

			\end{itemize}

\pagebreak

\item $(s_{1}, s_{2}) \tarrow{\lambda}{} (s'_{1}, s'_{2})$ iff one of the following conditions is fulfilled:

			\begin{itemize}

\item $s_{1} \tarrow{\mu}{1} s'_{1}$, $s_{2} \tarrow{\gamma}{2} s'_{2}$, $(s'_{1}, s'_{2}) = (s_{1},
s_{2})$, and $\lambda = \sum_{(s_{1}, \mu', s'_{1}) \in \! \subtarrow{}{1}} \mu' + \sum_{(s_{2}, \gamma',
s'_{2}) \in \! \subtarrow{}{2}} \gamma'$.

\item $s_{1} \tarrow{\mu}{1} s'_{1}$, $s'_{2} = s_{2}$, $s'_{1} \neq s_{1}$ or $s_{2} \hspace{0.3cm} /
\hspace{-0.5cm} \tarrow{}{2}$, $s_{2} \hspace{0.3cm} / \hspace{-0.5cm} \arrow{\tau}{2}$, and $\lambda =
\sum_{(s_{1}, \mu', s'_{1}) \in \! \subtarrow{}{1}} \mu'$.

\item $s_{2} \tarrow{\gamma}{2} s'_{2}$, $s'_{1} = s_{1}$, $s'_{2} \neq s_{2}$ or $s_{1} \hspace{0.3cm} /
\hspace{-0.5cm} \tarrow{}{1}$, $s_{1} \hspace{0.3cm} / \hspace{-0.5cm} \arrow{\tau}{1}$, and $\lambda =
\sum_{(s_{2}, \gamma', s'_{2}) \in \! \subtarrow{}{2}} \gamma'$.
\fullbox

			\end{itemize}

		\end{itemize}

	\end{definition}

Similar to~\cite{EHZ10}, from now on we uniformly treat the action transitions and the Markov timed
transitions of an MA $(S, A, \! \arrow{}{} \!, \! \tarrow{}{} \!)$ by considering time passage as a special
action $\chi$ and viewing the MA as a triple $(S, A^{\chi}, \! \arrow{}{} \!)$ where:

	\begin{itemize}\label{def:transf}

\item $A^{\chi} = A \cup \{ \chi(\lambda) \mid \lambda \in \realns_{\ge 0} \}$.

\item $s \arrow{\alpha}{} \Delta$ iff either $\alpha \in A$ with an identical action transition being
present in the original MA, or the following conditions are met:

		\begin{itemize}

\item $s \hspace{0.3cm} / \hspace{-0.5cm} \arrow{\tau}{}$.

\item $\alpha = \chi(\lambda)$ with $\lambda = \sum_{(s, \mu, s') \in \! \subtarrow{}{}} \mu$.

\item If $\lambda > 0$, then $\Delta(s') = \sum_{(s, \mu', s') \in \! \subtarrow{}{}} \mu' / \lambda$ for all
$s' \in S$, else $\Delta = \delta_{s}$.

		\end{itemize}

	\end{itemize}

\noindent
Notice that a state $s \in S$ can have at most one outgoing transition of the form $s
\arrow{\chi(\lambda)}{} \Delta$, with $\lambda$ and $\Delta$ being consistent with the race policy governing
$\! \tarrow{}{} \!$. This can be considered a probabilistic form of time determinism. In case of parallel
composition, it yields $(s_{1}, s_{2}) \arrow{\chi(\lambda)}{} \Delta$ iff one of the following holds:

	\begin{itemize}

\item $s_{1} \arrow{\chi(\mu)}{1} \Delta_{1}$, $s_{2} \arrow{\chi(\gamma)}{2} \Delta_{2}$, $\lambda = \mu +
\gamma$, and $\Delta = (\frac{\mu}{\lambda} \odot (\Delta_{1} \otimes \delta_{s_{2}})) \oplus
(\frac{\gamma}{\lambda} \odot (\delta_{s_{1}} \otimes \Delta_{2}))$.

\item $s_{1} \arrow{\chi(\lambda)}{1} \Delta_{1}$, $s_{2} \hspace{0.3cm} / \hspace{-0.5cm}
\arrow{\chi(\gamma)}{2}$ for all $\gamma \in \realns_{\ge 0}$, $s_{2} \hspace{0.3cm} / \hspace{-0.5cm}
\arrow{\tau}{2}$, and $\Delta = \Delta_{1} \otimes \delta_{s_{2}}$.

\item $s_{2} \arrow{\chi(\lambda)}{2} \Delta_{2}$, $s_{1} \hspace{0.3cm} / \hspace{-0.5cm}
\arrow{\chi(\mu)}{1}$ for all $\mu \in \realns_{\ge 0}$, $s_{1} \hspace{0.3cm} / \hspace{-0.5cm}
\arrow{\tau}{1}$, and $\Delta = \delta_{s_{1}} \otimes \Delta_{2}$.

	\end{itemize}

%
\subsection{Internal Transition Trees and Weak Transitions}
\label{sec:int_trans_tree_weak_trans}
%

The definition of weak bisimilarity for Markov automata given in~\cite{EHZ10} relies on \emph{labeled
trees}. For each such tree $\calt$, we denote by $\ms{nodes}(\calt)$ the set of its nodes, by
$\ms{leaves}(\calt)$ the set of its leaves, and by $\varepsilon_{\calt}$ its root. When $\calt$ only
contains $\varepsilon_{\calt}$, the node $\varepsilon_{\calt}$ is considered a leaf. Given $\sigma \in
\ms{nodes}(\calt)$, we denote by $\ms{children}(\sigma)$ the set of nodes reachable in one step from
$\sigma$.

For an MA $(S, A^{\chi}, \! \arrow{}{} \!)$, we use $(S \times \realns_{[0, 1]} \times (A^{\chi} \cup \{
\bot \}))$-labeled trees that somehow represent \emph{resolutions of nondeterminism}. Each node $\sigma$ is
labeled with the corresponding state $\ms{Sta}(\sigma) \in S$, the execution probability $\ms{Prob}(\sigma)
\in \realns_{[0, 1]}$ of the only path from $\varepsilon_{\calt}$ to $\sigma$, and the action
$\ms{Act}(\sigma) \in A^{\chi} \cup \{ \bot \}$ chosen to proceed.

	\begin{definition}\label{def:trans_tree_int}

Let $\calm = (S, A^{\chi}, \! \arrow{}{} \!)$ be an MA. A \emph{transition tree} $\calt$ for $\calm$ is an
$(S \times \realns_{[0, 1]} \times (A^{\chi} \cup \{ \bot \}))$-labeled tree that satisfies the following
conditions:

		\begin{enumerate}

\item $\ms{Prob}(\varepsilon_{\calt}) = 1$.

\item For each $\sigma \in \ms{leaves}(\calt)$, $\ms{Act}(\sigma) = \bot$.

\item For each $\sigma \in \ms{nodes}(\calt) \setminus \ms{leaves}(\calt)$, there is $\Delta$ such that
$\ms{Sta}(\sigma) \arrow{\ms{Act}(\sigma)}{} \Delta$ with:
\cws{10}{\hspace*{-0.6cm} \ms{Prob}(\sigma) \odot \Delta \: = \: \lsp (\ms{Sta}(\sigma'),
\ms{Prob}(\sigma')) \mid \sigma' \in \ms{children}(\sigma) \rsp}

		\end{enumerate}

\noindent
The distribution induced by $\calt$ on its leaves is defined as:
\cws{-2}{\Delta_{\calt} \: = \: \bigoplus_{\sigma \in \ms{leaves}(\calt)} \lsp (\ms{Sta}(\sigma),
\ms{Prob}(\sigma)) \rsp}
We say that $\calt$ is \emph{internal} iff, for each $\sigma \in \ms{nodes}(\calt)$, $\ms{Act}(\sigma) \in
\{ \tau, \bot \}$.
\fullbox

	\end{definition}

Given $s \in S$ and $\Delta \in \ms{Distr}(S)$, \emph{weak transitions} based on internal transition trees
and variants thereof are introduced as follows:

	\begin{itemize}

\item $s \warrow{}{} \Delta$ iff $\Delta$ is induced by an internal transition tree $\calt$ with
$\ms{Sta}(\varepsilon_{\calt}) = s$.

\item $s \warrow{\alpha}{} \Delta$ iff $\Delta$ is induced by a transition tree $\calt$ with
$\ms{Sta}(\varepsilon_{\calt}) = s$, such that along every maximal path from $\varepsilon_{\calt}$:

		\begin{itemize}

\item the action label of at least one inner node is $\alpha$ if $\alpha = \tau$,

\item the action label of precisely one inner node is $\alpha$ if $\alpha \neq \tau$,

		\end{itemize}

while the action label of all the other inner nodes is $\tau$.

\item $s \warrow{\hat{\alpha}}{} \Delta$ iff either $\alpha = \tau$ and $s \warrow{}{} \Delta$, or $\alpha
\neq \tau$ and $s \warrow{\alpha}{} \Delta$.

	\end{itemize}

\noindent
Requiring the target $\Delta$ of a weak transition to be a \emph{full} distribution ensures that all the
paths of the tree inducing the weak transition are of finite length, or that all of its infinite paths have
probability~$0$. In the first case, the tree is not necessarily finite, because some node may have countably
many children.

Convex combinations of identically labeled weak transitions are defined as follows: $s \warrow{\alpha}{\rm
c} \Delta$ iff there exist $n \in \natns_{\ge 1}$, $(p_{i} \in \realns_{]0, 1]} \mid 1 \le i \le n)$, and
$(s \warrow{\alpha}{} \Delta_{i} \mid 1 \le i \le n)$ such that $\sum_{1 \le i \le n} p_{i} = 1$ and $\Delta
= \bigoplus_{1 \le i \le n} p_{i} \odot \Delta_{i}$. Combined weak transition relations $\! \warrow{}{\rm c}
\!$ and $\! \warrow{\hat{\alpha}}{\rm c} \!$ are defined similarly.

%
\subsection{Strong and Weak Bisimilarities}
\label{sec:bisim}
%

Strong bisimilarity for Markov automata is a straightforward combination of strong bisimilarity for
probabilistic automata~\cite{Seg95a} and strong bisimilarity for interactive Markov chains~\cite{Her02}.

	\begin{definition}\label{def:strong_bisim}

Let $(S, A^{\chi}, \! \arrow{}{} \!)$ be an MA. An equivalence relation $\calb$ over $S$ is a \emph{strong
bisimulation} iff, whenever $(s_{1}, s_{2}) \in \calb$, then for all $\alpha \in A^{\chi}$ it holds that for
each $s_{1} \arrow{\alpha}{} \Delta_{1}$ there exists $s_{2} \arrow{\alpha}{} \Delta_{2}$ such that
$\Delta_{1}(C) = \Delta_{2}(C)$ for all $C \in S / \calb$. We write $s_{1} \sbis{} s_{2}$ to denote that
$(s_{1}, s_{2})$ is contained in some strong bisimulation.
\fullbox

	\end{definition}

The mix of the weak bisimilarities for the two classes of models is too fine for Markov automata.
In~\cite{EHZ10}, this drawback has been overcome by using combined weak transitions lifted to
\emph{sub}distributions. Each such $\alpha$-transition is obtained by weighting the target distribution of
the $\alpha$-transition from each source state with the probability assigned to that state by the source
subdistribution: $\Delta \warrow{\alpha}{\rm c} \Psi$ iff $s \warrow{\alpha}{\rm c} \Psi_{s}$ for all $s \in
\ms{supp}(\Delta)$ and $\Psi = \bigoplus_{s \in \ms{supp}(\Delta)} \Delta(s) \odot \Psi_{s}$. Combined weak
transition relations $\! \warrow{}{\rm c} \!$ and $\! \warrow{\hat{\alpha}}{\rm c} \!$ are lifted similarly.

	\begin{definition}\label{def:weak_bisim}

Let $(S, A^{\chi}, \! \arrow{}{} \!)$ be an MA. A relation $\calb$ over $\ms{Subdistr}(S)$ is a \emph{weak
bisimulation} iff, whenever $(\Delta_{1}, \Delta_{2}) \in \calb$, then $\ms{size}(\Delta_{1}) =
\ms{size}(\Delta_{2})$ and for all $\alpha \in A^{\chi}$ it holds that:

		\begin{itemize}

\item[(a)] For each $s_{1} \in \ms{supp}(\Delta_{1})$ there exist $\Delta'_{2}, \Delta''_{2} \in
\ms{Subdistr}(S)$ such that:

			\begin{enumerate}

\item $\Delta_{2} \warrow{}{\rm c} \Delta'_{2} \oplus \Delta''_{2}$ with $(\lsp (s_{1}, \Delta_{1}(s_{1}))
\rsp, \Delta'_{2}) \in \calb$ and $((\Delta_{1} \ominus s_{1}), \Delta''_{2}) \in \calb$.

\item For each $s_{1} \! \arrow{\alpha}{} \! \Psi_{1}$ there exists $\Delta'_{2} \warrow{\hat{\alpha}}{\rm
c} \Psi_{2}$ such that $(\Delta_{1}(s_{1}) \odot \Psi_{1}, \Psi_{2}) \! \in \! \calb$.

			\end{enumerate}

\item[(b)] Symmetric clause with the roles of $\Delta_{1}$ and $\Delta_{2}$ interchanged.

		\end{itemize}

\noindent
We write $\Delta_{1} \wbis{} \Delta_{2}$ to denote that $(\Delta_{1}, \Delta_{2})$ is contained in some weak
bisimulation. Moreover, we let $s_{1} \wbis{} s_{2}$ iff $\delta_{s_{1}} \wbis{} \delta_{s_{2}}$.
\fullbox

	\end{definition}

%
%
\section{Expected-Delay-Summing Weak Bisimilarity}
\label{sec:eds_weak_bisim}
%
%

In this section, we introduce a new weak bisimilarity~$\wbis{\rm eds}$ for Markov automata that, in addition
to abstracting from $\tau$-actions as~$\wbis{}$, sums up the expected values of consecutive exponentially
distributed delays possibly intertwined with $\tau$-actions. This is accomplished by relying on
\emph{reducible} projected transition trees. We prove that $\wbis{\rm eds}$ is an equivalence relation and
a congruence with respect to parallel composition, then we investigate the relationships between $\wbis{\rm
eds}$ and~$\wbis{}$.

%
\subsection{Projected Transition Trees: Components and Durations}
\label{sec:proj_trans_tree}
%

In general, a system is made out of several interacting sequential components. Therefore, we view a (global)
state of the MA at hand as a vector of local states. We denote by $\ms{Sta}(\sigma)[\ell]$ the state related
to sequential component~$\ell$ that occurs in the label of node $\sigma$ of a transition tree associated
with the MA. As shown in~\cite{Ber15} for exponentially timed actions, this component view is necessary to
achieve the congruence property with respect to parallel composition in the case of a weak bisimilarity that
adds up the expected values of exponentially distributed delays.

Furthermore, we extend transition trees by considering labels taken from the set $S \times \realns_{[0, 1]}
\times \realns_{\ge 0} \times (A^{\chi} \cup \{ \bot \})$, where each node $\sigma$ is additionally labeled
with the expected duration $\ms{Expd}(\sigma) \in \realns_{\ge 0}$ of the only path from the root
to~$\sigma$. To be precise, the states labeling the various nodes are \emph{global}. In contrast, to support
compositionality, the probability, the expected duration, and the action chosen to proceed labeling a node
are \emph{local} to the behavior of the considered component $\ell$ in isolation. Every path of such a tree
corresponds to a computation of component $\ell$ that is not forbidden by synchronization constraints or
maximal progress; the local transitions in that computation will be decorated with~$\ell$.

	\begin{definition}\label{def:proj_trans_tree}

Let $\calm = (S, A^{\chi}, \! \arrow{}{} \!)$ be an MA resulting from the parallel composition of $n \in
\natns_{\ge 1}$ MAs; we say that $\calm$ is \emph{sequential} when $n = 1$. Let $\ell \in \{ 1, \dots, n
\}$. An \emph{$\ell$-projected transition tree}~$\calt$ for~$\calm$ is an $(S \times \realns_{[0, 1]} \times
\realns_{\ge 0} \times (A^{\chi} \cup \{ \bot \}))$-labeled tree that satisfies the following conditions:

		\begin{enumerate}

\item $\ms{Prob}(\varepsilon_{\calt}) = 1$.

\item $\ms{Expd}(\varepsilon_{\calt}) = 0$.

\item For each $\sigma \in \ms{leaves}(\calt)$, $\ms{Act}(\sigma) = \bot$.

\item For each $\sigma \in \ms{nodes}(\calt) \setminus \ms{leaves}(\calt)$, there is $\Delta_{\ell}$ such
that the local transition $\ms{Sta}(\sigma)[\ell] \arrow{\ms{Act}(\sigma)}{\ell} \Delta_{\ell}$ contributes
to the derivation of some global transition $\ms{Sta}(\sigma) \arrow{\alpha}{} \Delta$ such that $\sigma'
\in \ms{children}(\sigma)$ iff $\ms{Sta}(\sigma') \in \ms{supp}(\Delta)$ with:
\cws{0}{\hspace*{-0.6cm} \ms{Prob}(\sigma) \odot \Delta_{\ell} \: = \: \lsp (\ms{Sta}(\sigma')[\ell],
\ms{Prob}(\sigma')) \mid \sigma' \in \ms{children}(\sigma) \rsp}
and for each $\sigma' \in \ms{children}(\sigma)$ it holds that:
\cws{8}{\hspace*{-0.6cm} \ms{Expd}(\sigma') \: = \: \left\{ \begin{array}{ll}
\ms{Expd}(\sigma) + \frac{1}{\lambda} & \hspace{0.5cm} \textrm{if $\ms{Act}(\sigma) = \chi(\lambda)$ and
$\lambda > 0$}
\\[0.1cm]
\ms{Expd}(\sigma) & \hspace{0.5cm} \textrm{if $\ms{Act}(\sigma) \in A \cup \{ \chi(0) \}$} \\
\end{array} \right.}
\fullbox

		\end{enumerate}

	\end{definition}

Variants of weak transitions based on internal $\ell$-projected transition trees are defined as in
Sect.~\ref{sec:int_trans_tree_weak_trans} and are respectively denoted by $\warrow{}{\ell}$,
$\warrow{\alpha}{\ell}$, and $\warrow{\hat{\alpha}}{\ell}$.

Since the leaves of $\calt$ are labeled with the expected duration of the corresponding paths from the root,
the distribution $\Delta_{\calt}$ induced by $\calt$ on its leaves can be decomposed into duration-indexed
subdistributions. To this aim, we define the subdistribution of $\Delta_{\calt}$ associated with $t \in
\realns_{\ge 0}$ as:
\cws{0}{\Delta_{\calt}^{t} \: = \: \bigoplus_{\sigma \in \{ \sigma' \in \ms{leaves}(\calt) \mid
\ms{Expd}(\sigma') = t \}} \lsp (\ms{Sta}(\sigma), \ms{Prob}(\sigma)) \rsp}
so that:
\cws{-2}{\Delta_{\calt} \: = \: \bigoplus_{t \in \{ \ms{Expd}(\sigma) \mid \sigma \in \ms{leaves}(\calt) \}}
\Delta_{\calt}^{t}.}
For the sake of convenience, we will often aggregate probabilities associated with leaves that are labeled
with the same expected duration by employing the notation:
\cws{0}{\Delta_{\calt} \: = \: \bigoplus_{i \in I} \Upsilon^{t_i}}
for some indexed set $\{ (t_i, \Upsilon^{t_i}) \}_{i \in I}$.

The above decomposition also extends to lifted combined weak transitions induced by internal variants of
$\ell$-projected transition trees. If $\Delta \warrow{\alpha}{\ell, \rm c} \Psi$, with $s
\warrow{\alpha}{\ell, \rm c} \Psi_{s}$ for all $s \in \ms{supp}(\Delta)$ and $\Psi = \bigoplus_{s \in
\ms{supp}(\Delta)} \Delta(s) \odot \Psi_{s}$, then, assuming $\Psi_{s} = \bigoplus_{i \in I_s}
\Psi_{s}^{t_i}$, we have that $\Psi = \bigoplus_{s \in \ms{supp}(\Delta)} \bigoplus_{i \in I_s} \Delta(s)
\odot \Psi_{s}^{t_i}$. \linebreak This can again be expressed in a more compact form as $\Psi = \bigoplus_{i
\in I} \Upsilon^{t_i}$ for some indexed set $\{ (t_i, \Upsilon^{t_i}) \}_{i \in I}$, provided that we join
all subdistributions $\Psi_{s}^{t_i}$ with the same $t_i$.

%
\subsection{Reducible Projected Transition Trees: Intertwining $\tau$ and $\chi(\lambda)$}
\label{sec:reduc_proj_trans_tree}
%

We now extend the notion of internal $\ell$-projected transition tree -- and add the corresponding weak
transitions -- by admitting nodes labeled with actions of the form $\chi(\lambda)$ that alternate with nodes
whose action label is $\tau$. The construction of this kind of tree proceeds as long as, in $\ell$, the
traversed local states have no alternative local transitions labeled with visible actions. In contrast, the
local states of $\ell$ contributing to the global states associated with the leaves cannot have local
transitions labeled with $\tau$ or $\chi(\lambda)$, unless they have a visible transition too. Following the
terminology of~\cite{Ber15}, we call the resulting tree reducible.

Taking into account alternative local transitions labeled with visible actions is crucial to achieve an
equivalence that does not abstract away from observable behaviors. For instance, should the context change
along the sequence of transitions between $s_{1}$ and $s_{6}$ of Fig.~\ref{fig:tau_exp_nondet_choice},
fusing those transitions into a unique Markov timed transition with the same expected duration would not be
appropriate. If there were an $a'$-transition from $s_{2}$ and a $b'$-transition from $s_{3}$, after an
exponentially distributed delay with rate~$\lambda$ we would notice a context change in the first MA that
cannot take place in the second one, thus preventing the transitions in the sequence from being merged.

	\begin{definition}\label{def:reduc_proj_trans_tree}

An $\ell$-projected transition tree $\calt$ for an MA $(S, A^{\chi}, \! \arrow{}{} \!)$ is \emph{reducible}
iff:

		\begin{enumerate}

\item For each $\sigma \in \ms{nodes}(\calt) \setminus \ms{leaves}(\calt)$, $\ms{Act}(\sigma) \in \{ \tau \}
\cup \{ \chi(\lambda) \mid \lambda \in \realns_{\ge 0} \}$.

\item For each $\sigma \in \ms{nodes}(\calt) \setminus \ms{leaves}(\calt)$, $\ms{Sta}(\sigma)[\ell]
\hspace{0.3cm} / \hspace{-0.5cm} \arrow{a}{\ell}$ for all $a \in A \setminus \{ \tau \}$.

\item For each $\sigma \in \ms{leaves}(\calt)$, $\ms{Sta}(\sigma)[\ell] \hspace{0.3cm} / \hspace{-0.5cm}
\arrow{\alpha}{\ell}$ for all $\alpha \in \{ \tau \} \cup \{ \chi(\lambda) \mid \lambda \in \realns_{\ge 0}
\}$ or $\ms{Sta}(\sigma)[\ell] \arrow{a}{\ell}$ for some $a \in A \setminus \{ \tau \}$.

\item Along every maximal path from $\varepsilon_{\calt}$, the action label of at least one inner node
belongs to $\{ \chi(\lambda) \mid \lambda \in \realns_{\ge 0} \}$.
\fullbox

		\end{enumerate}

	\end{definition}

Given $s \in S$ and $\Delta \in \ms{Distr}(S)$, we write $s \warrow{\chi}{\ell} \Delta$ iff $\Delta$ is
induced by a reducible $\ell$-projected transition tree $\calt$ with $\ms{Sta}(\varepsilon_{\calt}) = s$.
Combined and lifted variants of $\! \warrow{\chi}{\ell}$ are defined as usual.

	\begin{figure}[t]

\centerline{\includegraphics[width=4.7in]{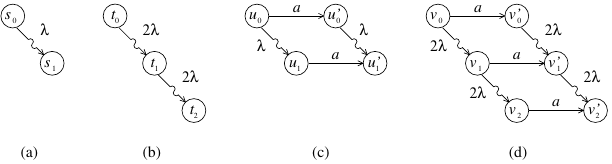}}
\caption{Reducing sequences of exponentially distributed delays}
\label{fig:reduc_seq_exp}

	\end{figure}

	\begin{example}\label{ex:seq_reduc_proj_trans_tree}

We start by considering systems made out of a single sequential component, for which we omit decoration
$\ell$ from transitions. In Fig.~\ref{fig:reduc_seq_exp}(a), we have $s_0 \warrow{\chi}{} \lsp (s_1, 1)
\rsp$, with the expected duration of the unique path being equal to $\frac{1}{\lambda}$. In
Fig.~\ref{fig:reduc_seq_exp}(b), we have $t_0 \warrow{\chi}{} \lsp (t_2, 1) \rsp$ with the expected duration
of the path leading from $t_0$ to $t_2$ being equal to $\frac{1}{2 \cdot \lambda} + \frac{1}{2 \cdot
\lambda} = \frac{1}{\lambda}$.

The two distributions induced by the two reducible projected transition trees will allow us to identify
$s_0$ and~$t_0$ by disregarding the intermediate state $t_1$. This would not be possible in the absence of
the third condition of Def.~\ref{def:reduc_proj_trans_tree}. In that case, we would also have $t_0
\warrow{\chi}{} \lsp (t_1, 1) \rsp$, whose associated expected duration $\frac{1}{2 \cdot \lambda}$ could
not be matched by any weak transition from $s_0$.

Also the second condition of Def.~\ref{def:reduc_proj_trans_tree}, which does not admit alternative visible
transitions along reducible computations, plays a role here. This is especially important for the
intermediate state $t_1$. \linebreak If this state had an alternative $a$-transition, in the absence of the
second condition $s_0$ and~$t_0$ would again be identified, in spite of the fact that an external observer
could see, at a certain point in time, the execution of a visible action only in the second system.
\fullbox

	\end{example}

	\begin{example}\label{ex:intro_reduc_proj_trans_tree}

Let us go back to Sect.~\ref{sec:intro}. The weak transition $s_1 \warrow{\chi}{} \lsp (s_6, 1) \rsp$ in
Fig.~\ref{fig:tau_exp_nondet_choice}(a) is induced by a reducible projected transition tree that associates
the expected duration $\frac{1}{\lambda} + \frac{1}{\mu} = \frac{\lambda + \mu}{\lambda \cdot \mu}$ with the
path from the root to the only leaf. This is the same expected duration associated with the weak transition
$t_1 \warrow{\chi}{} \lsp (t_2, 1) \rsp$, as well as the two occurrences of the weak transition $u_1
\warrow{\chi}{} \lsp (z_0, 1) \rsp$ induced by two alternative reducible projected transition trees deriving
from the nondeterministic choice at $u_3$. Therefore, we will be able to relate $s_0$, $t_0$, and $u_0$.

In Fig.~\ref{fig:tau_exp_nondet_choice}(b), we have the weak transition $s_1 \warrow{\chi}{} \lsp (u_2, p),
(v_2, 1 - p) \rsp$. The expected duration associated with the two leaves labeled with states $u_2$ and $v_2$
is $\frac{1}{\lambda} + \frac{1}{\mu} = \frac{\lambda + \mu}{\lambda \cdot \mu}$. This is also the expected
sojourn time associated with $t_1$, i.e., the inverse of $p \cdot \frac{\lambda \cdot \mu}{\lambda + \mu} +
(1 - p) \cdot \frac{\lambda \cdot \mu}{\lambda + \mu} = \frac{\lambda \cdot \mu}{\lambda + \mu}$. Thus, $t_1
\warrow{\chi}{} \lsp (w_0, p), \linebreak (z_0, 1 - p) \rsp$ will allow us to equate $s_0$ and~$t_0$.

In Fig.~\ref{fig:tau_exp_stoc_choice}, the reducible projected transition tree whose root is associated with
state $s_1$ has two leaves, labeled with states $u_1$ and $v_1$, respectively. By virtue of the race
condition at $s_3$, the probability of reaching such leaves is $\frac{\mu_1}{\mu_1 + \mu_2}$ and
$\frac{\mu_2}{\mu_1 + \mu_2}$, respectively, while the expected duration of both paths is $\frac{1}{\lambda}
+ \frac{1}{\mu_1 + \mu_2}$. Hence, we have $s_1 \warrow{\chi}{} \lsp (u_1, \frac{\mu_1}{\mu_1 + \mu_2}),
(v_1, \frac{\mu_2}{\mu_1 + \mu_2}) \rsp$. By anticipating the stochastic choice involving $\mu_1$
and~$\mu_2$, we obtain an analogous result, as $s_1' \warrow{\chi}{} \lsp (u_1', \frac{\mu_1}{\mu_1 +
\mu_2}), (v_1', \frac{\mu_2}{\mu_1 + \mu_2}) \rsp$ with the expected duration being as before.

Notice the analogy with the weak transition $t_1 \warrow{\chi}{} \lsp (w_0, \frac{\mu_1}{\mu_1 + \mu_2}),
(z_0, \frac{\mu_2}{\mu_1 + \mu_2}) \rsp$. By applying the race policy, we have that the expected sojourn
time in $t_1$ is again $\frac{1}{\lambda} + \frac{1}{\mu_1 + \mu_2}$, from which we derive that $s_0$,
$s'_0$, and~$t_0$ can be identified.
\fullbox

	\end{example}

	\begin{example}\label{ex:par_reduc_proj_trans_tree}

We now consider systems built from several components like the two MAs in Figs.~\ref{fig:reduc_seq_exp}(c)
and~(d), which are respectively obtained through the parallel composition of the two MAs in
Figs.~\ref{fig:reduc_seq_exp}(a) and~(b) with another MA that has only an $a$-transition. Each of the four
states of the MA in Fig.~\ref{fig:reduc_seq_exp}(c) can be paired with a state of the MA in
Fig.~\ref{fig:reduc_seq_exp}(d) as follows: $(u_0, v_0)$, $(u_1, v_2)$, $(u'_0, v'_0)$, $(u'_1, v'_2)$.

This is possible because the quantities associated with the nodes of reducible projected transition trees
are computed locally to the considered components. If the additional MA had a Markov timed transition with
rate $\mu$ instead of an $a$-transition, and the probabilities were computed globally, then the global
probability of going from $u_0$ to $u_1$ (which is $\frac{\lambda}{\lambda + \mu}$) and the global
probability of going from $v_0$ to~$v_2$ (which is $2 \cdot \frac{2 \cdot \lambda}{2 \cdot \lambda + \mu}$)
would be considered, which are different from each other.
\fullbox

	\end{example}

%
\subsection{A New Weak Bisimilarity: Definition and Properties}
\label{sec:eds_weak_bisim_def_prop}
%

We are finally in the position of defining a new weak bisimilarity for MAs that sums up expected values of
exponentially distributed delays while abstracting from $\tau$-actions. This is accomplished by considering
reducible projected transition trees \emph{in addition to} internal transition trees.

It is useful to extend to global states and distributions the notation $\! \arrow{}{\ell} \!$ for local
transitions employed in Defs.~\ref{def:proj_trans_tree} and~\ref{def:reduc_proj_trans_tree}. In the
following definition, we write $s \arrow{\chi(\lambda)}{\ell} \Delta$ to intend that $\Delta$ is induced by
an $\ell$-projected transition tree $\calt$ such that $\ms{Sta}(\varepsilon_{\calt}) = s$,
$\ms{Act}(\varepsilon_{\calt}) = \chi(\lambda)$, and $\ms{children}(\varepsilon_{\calt}) =
\ms{leaves}(\calt)$. Weak, combined, and lifted variants of the extended notation are as expected.

	\begin{definition}\label{def:eds_weak_bisim}

Let $(S, A^{\chi}, \! \arrow{}{} \!)$ be an MA. A relation $\calb$ over $\ms{Subdistr}(S)$ is an
\emph{expected-delay-summing weak bisimulation} iff, whenever $(\Delta_{1}, \Delta_{2}) \in \calb$, then
$\ms{size}(\Delta_{1}) = \ms{size}(\Delta_{2})$ and for all transition labels in $A^{\chi} \cup \{\chi\}$ it
holds that:

		\begin{itemize}

\item[(a)] For each $s_{1} \in \ms{supp}(\Delta_{1})$ there exist $\Delta'_{2}, \Delta''_{2} \in
\ms{Subdistr}(S)$ such that:

			\begin{enumerate}

\item $\Delta_{2} \warrow{}{\rm c} \Delta'_{2} \oplus \Delta''_{2}$ with $(\lsp (s_{1}, \Delta_{1}(s_{1}))
\rsp, \Delta'_{2}) \in \calb$ and $((\Delta_{1} \ominus s_{1}), \Delta''_{2}) \in \calb$.

\item For each $s_{1} \! \arrow{a}{} \! \Psi_{1}$ there exists $\Delta'_{2} \warrow{\hat{a}}{\rm c}
\Psi_{2}$ such that $(\Delta_{1}(s_{1}) \odot \Psi_{1}, \Psi_{2}) \! \in \! \calb$.

\item For each $s_{1} \arrow{\chi(\lambda)}{\ell_{1}} \Psi_{1}$ such that $s_{1} \hspace{0.3cm} /
\hspace{-0.5cm} \warrow{\chi}{\ell_{1}}$, there exists $\Delta'_{2} \warrow{\chi(\lambda)}{\ell_{2}, {\rm
c}} \Psi_{2}$ such that $(\Delta_{1}(s_{1}) \odot \Psi_{1}, \linebreak \Psi_{2}) \in \calb$.

\item For each $s_{1} \warrow{\chi}{\ell_{1}} \bigoplus_{i \in I} \Upsilon_{1}^{t_{i}}$ there exists
$\Delta'_{2} \warrow{\chi}{\ell_{2}, {\rm c}} \bigoplus_{i \in I} \Upsilon_{2}^{t_{i}}$ such that
$(\Delta_{1}(s_{1}) \odot \Upsilon_{1}^{t_{i}}, \Upsilon_{2}^{t_{i}}) \in \calb$ for all $i \in I$.

			\end{enumerate}

\item[(b)] Symmetric clause with the roles of $\Delta_{1}$ and $\Delta_{2}$ interchanged.

		\end{itemize}

\noindent
We write $\Delta_{1} \wbis{\rm eds} \Delta_{2}$ to denote that $(\Delta_{1}, \Delta_{2})$ is contained in
some expected-delay-summing weak bisimulation. Moreover, we let $s_{1} \wbis{\rm eds} s_{2}$ iff
$\delta_{s_{1}} \wbis{\rm eds} \delta_{s_{2}}$.
\fullbox

	\end{definition}

Condition~$1$ of Def.~\ref{def:eds_weak_bisim} coincides with condition~$1$ of Def.~\ref{def:weak_bisim},
while condition~$2$ of Def.~\ref{def:weak_bisim} is split into conditions~$2$ and~$3$ of
Def.~\ref{def:eds_weak_bisim}. In particular, by virtue of condition~$3$ above, Markov timed transitions
locally enabled by component $\ell_1$ are treated according to $\wbis{}$ whenever component $\ell_1$ does
not enable weak transitions induced by a \emph{reducible} $\ell_1$-projected transition tree. This is the
case when the Markov timed transition is locally alternative to a visible action transition, or in the
presence of time-divergence, i.e., an infinite sequence of Markov timed transitions, because the third
condition of Def.~\ref{def:reduc_proj_trans_tree} prevents reducible projected transition trees from being
generated as long as no state is encountered that enables a visible action or has no outgoing transitions.

On the other hand, condition~$4$ above states that every weak transition induced by a \emph{reducible}
\linebreak $\ell_1$-projected transition tree must be matched by a weak transition induced by a
\emph{reducible} $\ell_2$-projected transition tree, where $\ell_1$ is a component of the first process
while $\ell_2$ is a component of the second process. Notice that the target distributions are decomposed
into duration-indexed subdistributions prior to the application of the expected-delay-summing weak
bisimulation check, so to ensure that the matching also takes expected durations into account.

	\begin{example}\label{ex:eds_reduc_proj_trans_tree}

Following the discussions in Exs.~\ref{ex:seq_reduc_proj_trans_tree}, \ref{ex:intro_reduc_proj_trans_tree},
and~\ref{ex:par_reduc_proj_trans_tree}, we can establish that:

		\begin{itemize}

\item $s_0 \wbis{\rm eds} t_0 \wbis{\rm eds} u_0$ in Fig.~\ref{fig:tau_exp_nondet_choice}(a). For instance,
since $t_3 \wbis{\rm eds} z_1$, it is easy to verify that $\calb = \linebreak \{
(\delta_{t_{0}},\delta_{u_{0}}), (\delta_{t_{1}},\delta_{u_{1}}), (\delta_{t_{2}},\delta_{z_{0}}),
(\delta_{t_{3}},\delta_{z_{1}}) \}$ is an expected-delay-summing weak bisimulation. First, when applying
Def.~\ref{def:eds_weak_bisim}, for any pair $(\Delta_{1}, \Delta_{2}) \in \calb$ no splitting of
$\Delta_{2}$ is needed (i.e., $\Delta_{2} = \Delta'_{2}$). Then, $(\delta_{t_{2}},\delta_{z_{0}}) \in \calb$
as a consequence of condition~$2$ of Def.~\ref{def:eds_weak_bisim} and of the fact that $t_3 \wbis{\rm eds}
z_1$; $(\delta_{t_{1}},\delta_{u_{1}}) \in \calb$ follows by applying condition~$4$ of
Def.~\ref{def:eds_weak_bisim} (in particular, consider the weak transitions shown in
Ex.~\ref{ex:intro_reduc_proj_trans_tree}); $(\delta_{t_{0}},\delta_{u_{0}})$ follows by applying again
condition~$2$ of Def.~\ref{def:eds_weak_bisim}.

\item $s_0 \wbis{\rm eds} t_0$ in Fig.~\ref{fig:tau_exp_nondet_choice}(b).

\item $s_0 \wbis{\rm eds} t_0 \wbis{\rm eds} s_0'$ in Fig.~\ref{fig:tau_exp_stoc_choice}.

\item $s_{0} \wbis{\rm eds} t_{0}$ and $u_{0} \wbis{\rm eds} v_{0}$ in Fig.~\ref{fig:reduc_seq_exp}.  For
instance, consider the two MAs in Figs.~\ref{fig:reduc_seq_exp}(c) and~(d), which are obtained as discussed
in Ex.~\ref{ex:par_reduc_proj_trans_tree}. In particular, the former results from the parallel composition
of components $\ell_1$ and $\ell_3$, which are the MA of Fig.~\ref{fig:reduc_seq_exp}(a) and an MA that has
only an $a$-transition, respectively, while the latter from the parallel composition of components $\ell_2$,
which is the MA of Fig.~\ref{fig:reduc_seq_exp}(b), and $\ell_3$. Since $u'_1 \wbis{\rm eds} v'_2$, it holds
that $\calb = \{ (u_0, v_0), (u_1, v_2), (u'_0, v'_0), (u'_1, v'_2) \}$ is an expected-delay-summing weak
bisimulation. On one hand, $u_0 \warrow{\chi}{\ell_1} \lsp (u_1, 1) \rsp$, with the expected duration of the
unique path being equal to $\frac{1}{\lambda}$, while $v_0 \warrow{\chi}{\ell_2} \lsp (v_2, 1) \rsp$, with
the expected duration of the unique path being equal to $\frac{1}{2 \cdot \lambda} + \frac{1}{2 \cdot
\lambda} = \frac{1}{\lambda}$. Then, $u_1$ (resp., $v_2$) enables an $a$-transition leading to $u'_1$
(resp., $v'_2$). On the other hand, both $u_0$ and $v_0$ enable an $a$-transition, leading to $u'_0$ and
$v'_0$, respectively. Then, similarly as above, $u'_0 \warrow{\chi}{\ell_1} \lsp (u'_1, 1) \rsp$ is matched
by $v'_0 \warrow{\chi}{\ell_2} \lsp (v'_2, 1) \rsp$. Notice that $\calb$ does not include pairs containing
the intermediate states $v_1$ and $v'_1$. On one hand, this is necessary to match the weak transitions of
the two MAs. On the other hand, this is sufficient as, by virtue of the interleaving semantics of parallel
composition, the visible transitions enabled in $v_1$ (resp., $v'_1$) are the same as those enabled in $v_0$
and $v_2$ (resp., $v'_0$ and $v'_2$).
\fullbox

		\end{itemize}

	\end{example}

The relation $\! \wbis{\rm eds} \!$ turns out to be reflexive, symmetric, transitive, and substitutive with
respect to parallel composition.

	\begin{theorem}\label{thm:equivalence}

Let $\calm = (S, A^{\chi}, \! \arrow{}{} \!)$ be an MA. Then $\wbis{\rm eds}$ is an equivalence relation
over $\ms{Subdistr}(S)$.
\fullbox

	\end{theorem}

\begin{figure}[t]

\centerline{\includegraphics[width=3.3in]{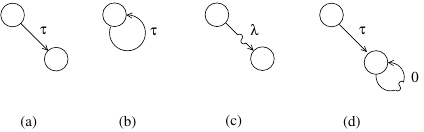}}
\caption{Interplay among maximal progress, $\tau$-divergence, and Markov timed transitions}
\label{fig:div_tau}

	\end{figure}

In order for the congruence property to hold, as pointed out in~\cite{SS14} it is essential to generate
\linebreak $\chi(0)$-selfloops for those states of the MAs at hand having neither $\tau$-transitions nor
Markov timed transitions. If such $\chi(0)$-selfloops were not generated by the transformation at the end of
Sect.~\ref{sec:ma}, then the interplay among maximal progress, $\tau$-divergence (i.e., an infinite sequence
of $\tau$-transitions), and Markov timed transitions would break compositionality. This can be seen by
considering a $\tau$-convergent MA with two states connected by a $\tau$-transition, see
Fig.~\ref{fig:div_tau}(a), and a $\tau$-divergent MA with a single state featuring a $\tau$-selfloop, see
Fig.~\ref{fig:div_tau}(b), each composed in parallel with the MA illustrated in Fig.~\ref{fig:div_tau}(c),
which has two states connected by a Markov timed transition. If no $\chi(0)$-selfloop were added to the
final state of the first MA, as shown in Fig.~\ref{fig:div_tau}(d), then the two MAs of
Figs.~\ref{fig:div_tau}(a) and~(b) would be identified, but the two composed MAs would be told apart.  In
fact, notice that the parallel composition of the MAs of Figs.~\ref{fig:div_tau}(a) and~(c) enables the
Markov timed transition after the execution of the $\tau$-transition, while the parallel composition of the
MAs of Figs.~\ref{fig:div_tau}(b) and~(c) executes $\tau$-transitions only. On the other hand, the two MAs
of Figs.~\ref{fig:div_tau}(a) and~(d) are not identified by $\wbis{\rm eds}$, as the Markov timed transition
of the latter cannot be matched by the former.

Analogous considerations on compositionality and sensitivity to $\tau$-divergence were also made in the IMC
setting of~\cite{Her02} and the pure Markovian setting of~\cite{Ber15}. In contrast, in the MA setting
of~\cite{DH13} the problem was circumvented by using a parallel composition operator that requires all
components to let time advance, in the same spirit as the deterministically timed model of~\cite{MT90,Yi91}.

	\begin{theorem}\label{thm:congruence}

Let $\calm_{k} = (S_{k}, A_{k}^{\chi}, \! \arrow{}{k} \!)$ be an MA for $k = 1, 2$. Let $A \subseteq (A_{1}
\cup A_{2}) \setminus \{ \tau \}$ and consider the parallel composition $\calm_{1} \pco{A} \calm_{2}$.  Let
$s_{1}, s'_{1} \in S_1$ and $s_{2} \in S_2$. If $s_{1} \wbis{\rm eds} s'_{1}$, then $(s_{1}, s_{2})
\wbis{\rm eds} (s'_{1}, s_{2})$.
\fullbox

	\end{theorem}

The following theorem states that $\wbis{\rm eds}$ is a conservative extension of $\wbis{}$ for sequential
MAs. These MAs constitute the common ground of the two equivalences, given that summing up expected delays
can be done compositionally only if the component structure is elicited. Investigating the relation between
the two equivalences in the case of generic MAs remains an open challenge unless renouncing to the
compositionality result of Thm.~\ref{thm:congruence}. Indeed, on one hand, condition~$4$ of
Def.~\ref{def:eds_weak_bisim} is critical to achieve the congruence property. On the other hand, it imposes
local conditions over the reducible behaviors of matching components that are completely ignored by
$\wbis{}$, as this equivalence abstracts away from the component-based structure of the system.

	\begin{theorem}\label{thm:cons_ext_sing_seq}

Let $(S, A^{\chi}, \! \arrow{}{} \!)$ be a sequential MA. Let $\Delta_{1}, \Delta_{2} \in \ms{Subdistr}(S)$.
If $\Delta_{1} \wbis{} \Delta_{2}$, then \linebreak $\Delta_{1} \wbis{\rm eds} \Delta_{2}$.
\fullbox

	\end{theorem}

%
%
\section{Application to Timed Labeled Transition Systems}
\label{sec:det_time}
%
%

So far, we have considered time passing described through \emph{exponentially distributed delays}, which is
typical of shared-resource systems. In this section, we consider a timed extension of labeled transition
systems inspired by~\cite{MT90,Yi91}, which is based on \emph{fixed delays} as in real-time systems.

	\begin{definition}\label{def:tlts}

A \emph{timed labeled transition system (TLTS)} is a tuple $(S, A, \! \arrow{}{} \!, \! \tarrow{}{} \!)$
where:

		\begin{itemize}

\item $S$ is a nonempty, possibly uncountable set of states.

\item $A$ is a set of actions containing at least the internal action $\tau$.

\item $\! \arrow{}{} \! \subseteq S \times A \times S$ is an action-transition relation.

\item $\! \tarrow{}{} \! \subseteq S \times \realns_{\ge 0} \times S$ is a time-transition relation such
that for all $s \in S$:

			\begin{itemize}

\item If $s \tarrow{0}{} s'$ for some $s' \in S$, then $s' = s$ \emph{(zero delay)}.

\item If $s \tarrow{t}{} s'_{1}$ and $s \tarrow{t}{} s'_{2}$ for some $s'_{1}, s'_{2} \in S$ and $t \in
\realns_{\ge 0}$, then $s'_{1} = s'_{2}$ \emph{(time determinism)}.

\item $s \tarrow{t_{1} + t_{2}}{} s''$ iff $s \tarrow{t_{1}}{} s'$ and $s' \tarrow{t_{2}}{} s''$ \emph{(time
additivity)}.

\item If $s \arrow{\tau}{} s'$ for some $s' \in S$, then $s \hspace{0.3cm} / \hspace{-0.5cm} \tarrow{}{}$
\emph{(maximal progress)}.
\fullbox

			\end{itemize}

		\end{itemize}

	\end{definition}

By analogy with MA, the passage of time $t \in \realns_{\ge 0}$ can be viewed as a special action that,
instead of $\varepsilon(t)$ as in~\cite{Yi91}, we denote by $\chi(t)$. Under this view, from now on we
consider a TLTS as a triple $(S, A^{\chi}, \! \arrow{}{} \!)$.

A notion of weak bisimilarity for TLTSs was studied in~\cite{Yi91,MT92}. It is essentially based on Milner's
\linebreak weak bisimilarity plus the capability of summing up fixed delays while abstracting from
$\tau$-actions. \linebreak Weak transitions are defined as follows:

	\begin{itemize}

\item $\! \warrow{}{} \! = (\! \arrow{\tau}{} \!)^{*}$.

\item $\! \warrow{a}{} \! = \! \warrow{}{} \!\! \arrow{a}{} \!\! \warrow{}{} \!$.

\vspace{0.1cm}
\item $\! \warrow{\hat{a}}{} \! = \! \left\{ \begin{array}{ll}
\warrow{}{} & \hspace{0.2cm} \textrm{if $a = \tau$} \\
\warrow{a}{} & \hspace{0.2cm} \textrm{if $a \neq \tau$} \\
\end{array} \right.$.

\vspace{0.1cm}
\item $\! \warrow{\chi(t)}{} \! = \! \warrow{}{} \!\! \arrow{\chi(t_{1})}{} \!\! \warrow{}{} \!\! \dots \!\!
\warrow{}{} \!\! \arrow{\chi(t_{n})}{} \!\! \warrow{}{} \!$ where $t = \sum_{1 \le i \le n} t_{i}$ for $n
\in \natns_{\ge 1}$.

	\end{itemize}

	\begin{definition}\label{def:bisim_tlts}

Let $(S, A^{\chi}, \! \arrow{}{} \!)$ be a TLTS. A symmetric relation $\calb$ over $S$ is a \emph{timed weak
bisimulation} iff, whenever $(s_{1}, s_{2}) \in \calb$, then for all actions $a \in A$ and amounts of time
$t \in \realns_{\ge 0}$ it holds that:

		\begin{itemize}

\item For each $s_{1} \arrow{a}{} s'_{1}$ there exists $s_{2} \warrow{\hat{a}}{} s'_{2}$ such that $(s'_{1},
s'_{2}) \in \calb$.

\item For each $s_{1} \arrow{\chi(t)}{} s'_{1}$ there exists $s_{2} \warrow{\chi(t)}{} s'_{2}$ such that
$(s'_{1}, s'_{2}) \in \calb$.

		\end{itemize}

\noindent
We write $s_{1} \wbis{\rm t} s_{2}$ to denote that $(s_{1}, s_{2})$ is contained in some timed weak
bisimulation.
\fullbox

	\end{definition}

Relations $\wbis{\rm eds}$ and $\wbis{\rm t}$ share the idea of summing up delays while abstracting from
$\tau$-actions. \linebreak The theorem below performs a more precise comparison based on an adaptation of
the $\wbis{\rm eds}$ construction to TLTS models suitably modified according to the following
considerations:

	\begin{itemize}

\item We regard TLTS transitions as leading to Dirac distributions over states.

\item We assume that actions can only be instantaneously (as opposed to continuously) enabled.

\item We assume that all $\chi(t)$-transitions have $t \in \realns_{> 0}$ and are expressed up to time
decomposability (an aspect of time additivity), in the sense that a transition of the form $s 
\arrow{\chi(t)}{} s'$ with $t \in \realns_{> 0}$ subsumes all the possible computations from $s$ to $s'$ 
whose total duration is $t$, such that each of the intermediate states at time distance $t' \in 
\realns_{]0, t[}$ from $s$ does not enable any action and has only computations to $s'$ of total 
duration $t - t'$. Due to time determinism, this guarantees that every TLTS state will have at most one 
outgoing transition labeled with $\chi(t)$, where $t \in \realns_{> 0}$.

\item When building transition trees, the expected duration of the only path from the root to a node is
computed as the sum of the time delays -- rather than their inverses as in the fourth condition of
Def.~\ref{def:proj_trans_tree} -- occurring along that path. The reason is that, in a deterministically
timed setting, a delay coincides with its expected value.

	\end{itemize}


	\begin{theorem}\label{thm:cons_ext_timed}

Let $(S, A^{\chi}, \! \arrow{}{} \!)$ be a modified TLTS originated from a system made out of a single
\linebreak sequential component, whose states have at most one outgoing $\tau$-transition each. Let $s_{1},
s_{2} \in S$. \linebreak If $s_{1} \wbis{\rm eds} s_{2}$, then $s_{1} \wbis{\rm t} s_2$.
\fullbox

	\end{theorem}

The result does not hold in the presence of choices among $\tau$-transitions, because $\wbis{\rm t}$ is more
sensitive to them than $\wbis{\rm eds}$. For example, given $a,b \in A \setminus \{ \tau \}$, the systems
described in process algebraic style as $\chi(t_{1} + t_{2}) \, . \, (\tau \, . \, a + \tau \, . \, b)$ and
$\chi(t_{1}) \, . \, (\tau \, . \, \chi(t_{2}) \, . \, a + \tau \, . \, \chi(t_{2}) \, . \, b)$ are
identified by~$\wbis{\rm eds}$, but distinguished by $\wbis{\rm t}$. Intuitively, there is no reason to
distinguish them on the basis of the instant of time in which the internal choice is solved, as in both
cases an external observer should wait $t_{1} + t_{2}$ time units before interacting with the system. 

Moreover, the above implication cannot be reversed. For instance, given $a \in A \setminus \{ \tau \}$, the
process $a + \chi(t_{1} + t_{2})$ and the process $a + \chi(t_{1}) \, . \, \chi(t_{2})$ are identified
by~$\wbis{\rm t}$, but distinguished by $\wbis{\rm eds}$. The reason is that the latter cannot sum up delays
in the presence of \emph{locally} alternative visible actions.


%
%
\section{Conclusions}
\label{sec:concl}
%
%

Building on previous work~\cite{EHZ10,Ber15}, we have incrementally extended the identification power of a
weak semantics for MAs by defining a new weak bisimulation congruence $\wbis{\rm eds}$ that, in addition
to abstracting from $\tau$-actions, sums up the expected values of consecutive exponentially distributed
delays possibly intertwined with $\tau$-actions. From an application viewpoint, $\wbis{\rm eds}$ can thus
serve as the semantical basis for state space reduction techniques more aggressive than those recently
developed for MAs in~\cite{TPS13}.

The relation $\wbis{\rm eds}$ has also been compared with the weak bisimilarity defined
in~\cite{Yi91,MT92} for TLTSs; in these models, it is standard to be capable of adding up expected delays
interleaved with $\tau$-actions. Therefore, the definition of $\wbis{\rm eds}$ constitutes a step towards
reconciling the semantics for stochastic time and deterministic time, a subject recently addressed
in~\cite{MDBD12}.

As far as future work is concerned, we plan to investigate equational and logical characterizations
of~$\wbis{\rm eds}$. Furthermore, since we have priviledged the achievement of the congruence property with
respect to the tradeoff emerged in~\cite{Ber15}, we intend to examine the preservation of quantitative
properties. This has been addressed in~\cite{Ber15} for stationary-state reward-based performance measures
in the case of pure Markovian models. An analogous result in the case of MAs needs to take into account
nondeterminism and hence the fact that only maximum and minimum values of performance measures can be
computed after applying suitable schedulers~\cite{GHHKT13}. Finally, we would like to adapt $\wbis{\rm eds}$
to a probabilistic extension of the TLTS model, in which the target of an action transition can be a general
probability distribution over states as in the MA model, so to study the relationships with the weak
bisimilarity defined in~\cite{LMT10}.

\medskip
\noindent
\textbf{Acknowledgment}: This work has been funded by MIUR-PRIN project CINA.

\bibliographystyle{eptcs}
\bibliography{qapl2015}

\end{document}